\documentclass[12pt,preprint]{aastex}
\usepackage{graphicx}
\newcommand{\Chandra} {{\em Chandra}}

\begin{document}

\slugcomment{Accepted for \Chandra\ Carina Complex Project ApJS Special Issue}
\shortauthors{Feigelson et al.}
\shorttitle{X-ray Star Clusters in Carina}

\title{X-ray Star Clusters in the Carina Complex}

\author{Eric D. Feigelson\altaffilmark{*}\altaffilmark{1}, Konstantin V. Getman\altaffilmark{1}, Leisa K. Townsley\altaffilmark{1}, Patrick S. Broos\altaffilmark{1}, Matthew S. Povich\altaffilmark{1,2},  Gordon P. Garmire\altaffilmark{1}, Robert R. King\altaffilmark{3}, Thierry Montmerle\altaffilmark{4}, Thomas Preibisch\altaffilmark{5}, Nathan Smith\altaffilmark{6,7}, Keivan G. Stassun\altaffilmark{8,9}, Junfeng Wang\altaffilmark{10}, Scott Wolk\altaffilmark{10}, Hans Zinnecker\altaffilmark{11}}

\altaffiltext{*} {edf@astro.psu.edu} 
\altaffiltext{1}{Department of Astronomy \& Astrophysics, Pennsylvania State University, 525 Davey Lab, University Park, PA 16802}
\altaffiltext{2}{NSF Postdoctoral Fellow}
\altaffiltext{3}{Astrophysics Group, College of Engineering, Mathematics and Physical Sciences, University of Exeter, Exeter EX4 4QL, UK}
\altaffiltext{4}{Institut d'Astrophysique de Paris, 98bis Boulevard Arago, F-75014 Paris, France}
\altaffiltext{5}{Universit\"ats-Sternwarte M\"unchen, Ludwig-Maximilians-Universit\"at, Scheinerstr.~1, 81679 M\"unchen, Germany} 
\altaffiltext{6}{Department of Astronomy, University of California, Berkeley, CA 94720}
\altaffiltext{7}{Steward Observatory, University of Arizona, 933 North Cherry Avenue, Tucson, AZ 85721, USA }
\altaffiltext{8}{Department of Physics \& Astronomy, Vanderbilt University, 6301 Stevenson Center Lane, Nashville, TN 37235}
\altaffiltext{9}{Department of Physics, Fisk University, 1000 17th Ave. N., Nashville, TN 37208}
\altaffiltext{10}{Harvard-Smithsonian Center for Astrophysics, 60 Garden Street, Cambridge, MA 02138}
\altaffiltext{11}{Astrophysical Institute Potsdam, An der Sternwarte 16, 14482 Potsdam, Germany}

\begin{abstract}

The distribution of young stars found in the {\it Chandra Carina Complex Project} (CCCP) is examined for clustering structure.  X-ray surveys are advantageous for identifying young stellar populations compared to optical and infrared surveys in suffering less contamination from nebular emission and Galactic field stars. The analysis is based on smoothed maps of a spatially complete subsample of $\sim 3000$  brighter X-ray sources classified as Carina members, and $\sim 10,000$ stars from the full CCCP sample.   The principal known clusters are recovered, and some additional smaller groups are identified. No rich embedded clusters are present, although a number of sparse groups are found. 

The CCCP reveals considerable complexity in clustering properties.  The Trumpler~14 and 15 clusters have rich stellar populations in unimodal, centrally concentrated structures several parsecs across.  Non-spherical internal structure is seen,  and large-scale low surface density distributions surround these rich clusters. Trumpler~16, in contrast, is comprised of several smaller clusters within a circular boundary.  Collinder~228 is a third type of cluster which extends over tens of parsecs with many sparse compact groups likely arising from triggered star formation processes. A widely dispersed, but highly populous, distribution of X-ray stars across the $\sim 50$~pc CCCP mosaic supports a model of past generations of star formation in the region.  Collinder~234, a group of massive stars without an associated cluster of pre-main sequence stars, may be part of this dispersed population.

\end{abstract}


\section{Introduction}

The Carina Nebula (NGC~3372) is perhaps the richest and most violent star forming complex in the Galaxy within several kiloparsecs of the Sun.  Its prominence arises because it has produced, and is now illuminated and excited by, a string of rich OB clusters, a `cluster of clusters'.  It is thus the nearest analog to extragalactic HII regions and starburst phenomena.  Most studies of its stellar population have concentrated on its massive stars: main sequence OB stars, supergiant O and Wolf-Rayet stars, and the eruptive luminous blue variable $\eta$ Car.  However, it is also important to study its lower mass population as both early cluster formation mechanisms and later dynamical evolution can cause spatial segregation by stellar mass.  The Initial Mass Function is also important as a tracer of star formation processes.  

The present study describes new findings on the spatial clustering of young stars in the Carina Nebula as derived from the new Chandra Carina Complex Project (CCCP), a 1.42~deg$^2$ mosaic obtained in the X-ray band with NASA's {\it Chandra X-ray Observatory} \citep{Townsley11a}.   \Chandra\ X-ray surveys are remarkably efficient at identifying large populations of both OB and lower mass pre-main sequence (PMS) stars in the rich young clusters lying 1-3~kpc distant \citep[see review by][]{Feigelson07}.  X-ray surveys suffer only mild contamination by older Galactic field stars \citep{Getman11} due to the rapid decline in magnetic flare activity after stars reach the main sequence.  X-ray selected populations nicely complement optical and infrared selected populations which, due to heavy contamination by field stars, are often restricted to lightly obscured and disk-bearing stars.  In the Carina region, the CCCP has located 14,368 X-ray sources of which 10,728 are classified as probable stellar members of the region \citep{Broos11a, Broos11b}.  
 
Due to complex spatial distributions of both cloud obscuration and nebular emission, the clustering structure of the Carina Nebula has not been clearly delineated.  The historical clusters were defined by Trumpler and Collinder in the 1930s from photographic plates.  In the 1970s, different conclusions were reached using photometrically selected OB stars: \citet{The71} find that Trumpler~14, Trumpler~15, and Trumpler~16 are distinct clusters with Trumpler~15 lying in front of the nebula; \citet{Feinstein73} find that Trumpler~14, Trumpler~16, and Collinder~232 are a contiguous large cluster.  \citet{Walborn73, Walborn95} infers that three clusters are present: Trumpler~16 and Collinder~228 are a single cluster divided by a dust lane, Trumpler~14 and Collinder~232 form a single cluster at a somewhat greater distance, and Trumpler~15 is a minor background cluster.    Photometric studies found that Bochum~10, Bochum~11, and Collinder~228 are distinct sparse open clusters \citep{Patat01, Carraro01b}.  CCD photometry led \citet{Tapia03} to conclude that Collinder~232 is not a distinct cluster, while \citet{Carraro04} concluded that  Collinder~232 is a physical aggregate.  In a recent review, \citet{Smith08} infer that Collinder~228 is a part of Trumpler~16.  A compact, semi-embedded cluster known as the Treasure Chest produces a bright mid-infrared source south of Trumpler~16, and a number of other fainter embedded star formation regions are known \citep{Smith05, Smith08}.   Part of the difficulty in delineating physical structures in Carina is that a Galaxy spiral arm is seen in projection along the line of sight. 

The lack of consensus on the cluster structure in the Carina Nebula is due to several types of confusion that optical and near-infrared surveys must treat: foreground and background star distributions (including OB stars unrelated to Carina along the line of sight), spatially varying dust obscuration and emission line nebulosity, and possibly spatially varying dust extinction laws within the Nebula.  X-ray surveys are mostly free from these problems, although they encounter other limitations.  Once contaminant sources are removed \citep{Getman11}, an X-ray survey performed with sufficiently high resolution optics will detect most OB stars and a large population of PMS stars down to a mass limit that is roughly dependent on the X-ray flux sensitivity limit \citep[see][for the statistical relation between X-ray luminosity and PMS mass]{Telleschi07}.  X-ray emission is strong at all stages of PMS evolution from Class I protostars through arrival onto the main sequence \citep{Preibisch05}, although luminosities decrease somewhat in systems with high accretion rates and sensitivities are reduced in regions of high obscuration.  Overall, X-ray surveys can be very effective in establishing the spatial structure of complex clustered environments such as the Carina Nebula.  Studies of this type have been carried out for the NGC 2024, NGC 2264, M~17, Rosette Nebula, W~3, NGC~6334, and other rich star forming complexes \citep[e.g.,][]{Skinner03, Flaccomio06, Broos07, Feigelson08, Feigelson09, Wang09}.

After a brief description of the CCCP dataset (\S~\ref{methods.sec}) and analysis methods, this paper presents an unbiased map and list of major young stellar clusters in the Carina Nebula region (\S~\ref{majclus.sec}).  A less-complete list of small star groups (\S~\ref{groups.sec}), some embedded in molecular clouds, and large-scale star density enhancements (\S~\ref{regions.sec}) is then developed.  We conclude with a list of star members in the clustered structures (\S~\ref{clusmem.sec}) and a clarification of the historical controversies outlined above (\S~\ref{disc.sec}).

\section{Observations and Methods \label{methods.sec}}

The CCCP observations and their analysis are described in detail by \citet{Townsley11a} and \citet{Broos11a}.  Twenty-two overlapping pointings with the Advanced CCD Imaging Spectrometer (ACIS), each subtending $17\arcmin \times 17\arcmin$, were observed.  They cover a 1.4~square degree region roughly defined by $10^h40^m < \alpha < 10^h50^m$ and $-59^\circ00\arcmin < \delta < -60^\circ30\arcmin$.  Typical exposures are 60~ks in duration, but the exposure map shows considerable spatial variation.  

Data analysis followed procedures described by \citet{Broos11a}.  After data cleaning, several procedures are used to scan the images for faint source candidates.  The final source list is iteratively selected based on a Poisson probability criterion with respect to the local background level.  The X-ray data are aligned to the 2MASS/Hipparcos frame and, in most cases, source positions have accuracies better than 0.5\arcsec.  Infrared counterparts are found by positional coincidences with 2MASS, VLT HAWK-I \citep{Preibisch11}, and other published catalogs.  Three classes of contamination --- foreground stars, background stars, and extragalactic sources --- are present \citep{Getman11}; \citet{Broos11b} assign class probabilities to individual CCCP sources.

The result of this analysis procedure is the identification of 14,368 X-ray sources, most very faint.  CCCP sources are designated CXOGNC for `\Chandra\ X-ray Observatory Great Nebula in Carina'.  The present study is restricted to the 10,728 sources classified as `probable Carina members', which we call the `full sample' .  However, due to telescope effects (off-axis mirror vignetting and degradation of the point spread function) and survey construction (overlapping fields with different exposures), factors of $\la 6$ spatial variations in sensitivity are present in the sample \citep[][\S~6.1]{Broos11a}. The result is an artificial clustering of sources towards the centers of each ACIS field.  To avoid confusing this instrumental clustering and intrinsic star clustering in the Carina region, we construct a subsample of 3,220 sources within the full sample defined by a threshold on observed X-ray photon flux, $\log F_{t,photon} > -5.9$ photon~s$^{-1}$~cm$^{-2}$ in the `total' ($0.5-8$~keV) \Chandra\ band \footnote{ This photon flux is defined by \citet[][\S7.4]{Broos10} using three measured total-band quantities as $F_{t,photon} = NetCounts\_t /  ExposureTimeNominal/MeanEffectiveArea\_t $.}. The entire survey area has been observed to this sensitivity level, and we call this subsample the `spatially complete sample'.  

Our science goals here are to define star groupings in the spatial distribution of \Chandra\ Carina members.  The statistical problem of unsupervised clustering of an inhomogeneous and multiscale spatial point process has no definitive solution.  (`Unsupervised' means that no prior knowledge of cluster number or location is utilized.) A variety of estimation methods can be used including agglomerative hierarchical clustering (e.g., the `friends-of-friends' algorithm), $k$-means partitioning, and data mining classification \citep{Duda00, Kaufman05}.  Due to the hierarchical complexity of the X-ray source distribution in the Carina Nebula, we found that traditional methods do not readily provide reliable results; for example, there is no single level at which the agglomerative clustering dendrogram can be cut that reveals all of the clearly evident clusters.   Density-dependent clustering algorithms such as {\it DBSCAN} may treat this problem \citep{Ester96, Pei09}, but we opt here for a simpler alternative approach of smoothing using kernel density estimation with normal (Gaussian) kernels \citep{Silverman92}.  We examined whether results would be different if the density estimation were performed in three dimensions---right ascension, declination, and median X-ray event energy representing absorption---but we found that few structures cover a wide range of absorption and no overlap between lightly and heavily obscured clusters was present.  Therefore, a two-dimensional treatment seems adequate.  

Kernel density estimation is a well-established technique to smooth individual data point locations into a continuous spatial distribution; that is, we convert the individual source $(\alpha,\delta)$ locations to a continuous map of source surface density.  Two parameters must be provided: the bandwidth of the kernel, and an appropriate density threshold that separates clusters from unclustered sources.  In the analysis below, we choose these parameters subjectively.  Therefore, the cluster lists and memberships provided here cannot be considered to be unique or validated to some statistical level of significance. There is no assurance that these clusters are physically real or fully represent the spatial structure of the Carina X-ray source population.  However, the method does have several advantages: no prior assumptions were made on the number, locations, or shapes of clusters; the entire dataset is treated in a uniform fashion; and the astronomical results can be clearly visualized and tabulated.  {\citet{Schmeja11} has compared several methods for finding simulated star clusters under realistic conditions and reports that binned star counts (a simple method similar to kernel density estimation) performs comparably or better than nearest neighbor, Voronoi tessellation, and pruned minimal spanning tree methods.  We note that a mathematical result known as the `No Free Lunch Theorem' states that there is no general way to establish that one clustering solution is better than another in a multivariate unsupervised classification problem \citep{Wolpert97}. 

The kernel density estimation was performed with the program {\it bkde2D} (2-dimensional binned kernel density estimator) in the {\it KernSmooth} package of the {\bf R} statistical programming environment.  {\bf R} is the largest public-domain statistical computing package\footnote{\url{http://r-project.org}};  see, for example, \citet{Crawley07} regarding its use.  The data were transformed from astronomical $(\alpha,\delta)$ coordinates to projected angular distances prior to smoothing to avoid a $\sim 3$\% distortion in the metric due to the spherical geometry across the CCCP mosaic.  The smoothed maps were then transformed back to right ascension and declination units for conve nient display.   Other analyses and displays were made using {\it Interactive Data Language}\footnote{\url{http://www.ittvis.com/idl}} and {\it ds9}\footnote{\url{hea-www.harvard.edu/RD/ds9/}}.

\section{Clustered Structures in Carina} 

\subsection{Principal X-ray Selected Star Clusters \label{majclus.sec}}

As outlined in the previous section, we cannot consider the full CCCP sample of 14,368 X-ray sources in the Carina region for two reasons: a fraction of the sources are foreground or background contaminants, and the survey has inhomogeneous sensitivity due to characteristics of the \Chandra\ telescope and the survey design.   Both of these effects are greatly reduced in the spatially complete sample of 3,220 probable Carina members above a photon flux limit described above.  

Figure~\ref{Complete_smooth.fig} shows the distribution of this spatially complete Carina member sample.  Panel (a) shows the individual star positions, while panel (b) shows the smoothed source surface density using a normal kernel with standard deviation $\sigma = 30$\arcsec.  Assuming a distance of 2.3~kpc to the Carina Nebula, this corresponds to a Gaussian kernel with $\sigma \simeq 0.3$~pc or full-width-half-maximum ${\rm FWHM}=0.8$~pc.   We select the density threshold for cluster identification to be the third contour (0.003 in the normalized convolution of the kernel with the data) in this surface density map.  The 20 surface density peaks above this threshold are labeled in the figure and listed in Table~1 as the principal X-ray selected clusters in the region.  We adopt the nomenclature `CCCP-Cl' for `\Chandra\ Carina Complex Project Cluster' to identify these clusters (column 1). The cluster positions (columns $2-3$) represent the surface density peaks; due to nonspherical distributions, these may not correspond to mean or median positions of the member stars.  When a larger $\sigma=40$\arcsec\/ kernel is used, the Tr~16 substructure becomes less distinct, and when a smaller $\sigma=20$\arcsec\/ is used, some of the small groups (\S~\ref{groups.sec}) begin to emerge.

While the spatially complete sample (3,220 bright probable Carina members) allows a spatially unbiased identification of rich clusters, we use the full sample (10,728 probable Carina members) to estimate properties of each cluster because the CCCP survey sensitivity variations are generally small on intra-cluster scales.  Figure~\ref{Full_maps.fig} shows the distribution of the full sample superposed on the contours of clusters found from Figure~\ref{Complete_smooth.fig}.   The figure also outlines the individual ACIS fields; they give a clear warning that the spatial variations in the full sample can be due to instrumental effects (off-axis sensitivity degradation and overlapping exposures) as well as to intrinsic star clustering in the Carina complex.  f Our use of a $\sigma=30$\arcsec\/ (0.3~pc) Gaussian kernel and omission of fainter X-ray sources also blurs substructure within the rich Tr~14, 15 and 16 clusters.  More complete analyses of the structure of these clusters are given by \citet{Ascenso07}, \citet{Wang11}, and \citet{Wolk11}, respectively.

To estimate the star population $N$ of each cluster, we give the number of sources in the spatially complete sample within some boundary around the cluster peak locations to be cluster members.   The contours, counted from the lowest contour in Figure~\ref{Complete_smooth.fig}b, are listed in column 4 and $N$ is given in column 5 of Table~1.  For example, contour ``2m'' represents the second lowest contour where ``m'' is appended when the contour was modified to produce a closed boundary.  Observed cluster populations range from $N \simeq 10$ to over 1000 X-ray selected stars.  

In PMS populations, useful estimates of line-of-sight interstellar gas column densities, $N_H$, can be obtained from the observed median X-ray event energy statistic \citep{Feigelson05, Getman10}.  Table~1 reports the average and standard deviation of this statistic, $\langle MedE\rangle$ (column 7), for the $N_{ME}$ (column 6) cluster members for which median event energy is available (sources with $>$4 net counts extracted). Visual absorption, $A_V$, can be estimated from the gas column densities, $N_H$, inferred from $\langle MedE\rangle$ assuming standard gas-to-dust ratios \citep{Vuong03}.  These $A_V$ estimates, ranging from $A_V \simeq 2$ to 10 magnitudes, are reported with low precision in column 8.

Correspondences to previously known optical and near-infrared star clusters (column 9) are established visually from maps given by \citet{Feinstein95} and by \citet{Smith08}. Note that Trumpler~16 is divided into seven clusters in our analysis.  Associations with mid-infrared clusters in the South Pillars region are based on the {\em Spitzer} IRAC study of \citet{Smith10}.  A $\star$ symbol indicates whether an X-ray cluster lies in the region occupied by the South Pillars and the large Collinder~228 star cluster discussed in \S~\ref{Collinder.sec}.   The final column of Table~1 lists the dominant massive stars within or near each cluster with spectral types obtained from \citet{Gagne11} or SIMBAD.  These are labelled $a$, $b$, $\ldots$, $y$ to identify them in Figure~\ref{Full_maps.fig} where they are plotted as triangles.  These are far from a complete list of OB stars, but give a general indication of the cluster richness in the region.  The X-ray OB population in the CCCP,  including their multiplicity and X-ray properties,  is discussed in detail by Gagn\'e et al. and \citet{Povich11a}.

\subsection{Small Groups of X-ray Stars} \label{groups.sec}

While the smoothed map of Figure~\ref{Complete_smooth.fig} is effective in finding clusters with scales around $0.5\arcmin-5\arcmin$\/ ($0.3-3$ pc) in a spatially unbiased fashion, it misses some small groups that are evident in visual examination of the full sample.  We therefore constructed  a smoothed density map of the full sample with a 10\arcsec\/ Gaussian kernel, and made a list of structures above a subjectively chosen density threshold.  This new list of small groups is restricted to regions lying outside the boundaries of the principal clusters defined in column (4) of Table 1.  Note that, due to instrumental spatial sensitivity variations across the field using the full sample (\S~2),  the search for sparse groups cannot be considered complete to some fixed flux limit. The group list also depends on the choice of smoothing kernel; a larger 15\arcsec\/ kernel merges some of the closer groups into larger structures.

The resulting list of 31 small X-ray selected groups of probable Carina members is presented in Table 2; columns 4-7 have the definitions as in Table 1.  The group member tallies, $N$, are measured in 40\arcsec\/  diameter circular regions centered on the local peaks of the smoothed map.   These groups are designated `CCCP-Gp' to distinguish them from the larger and richer CCCP-Cl clusters.  The groups are both compact and sparse, exhibiting only $5 < N < 12$ X-ray members.  Typical member surface densities are $50-100$ stars pc$^{-2}$.  Most lie in the southern portion of the mosaic, and many have counterparts among the mid-infrared clusters identified by \citet{Smith10} from {\em Spitzer} IRAC images of the South Pillars region.  The early-type stars in this region are often considered to be members of an extended open cluster Collinder~228 \citep{Feinstein76}, so one might choose to associate these small X-ray groups with Collinder~228 (\S~\ref{Collinder.sec}).

\subsection{Large-Scale Regions with Young Stars \label{regions.sec}}

The clusters and groups defined in Tables 1 and 2 contain $\sim 3,000$ and $\sim 300$ stars in the full sample, respectively.  But more than twice as many stars lie outside these cluster and group boundaries in large regions with enhanced stellar surface densities.  We have divided these unclustered stars into three spatial regions shown in Figure~\ref{Complete_smooth.fig}b:  Region A  around and between Trumpler~14 and 15; Region B within and south of the Trumpler~16 complex; and Region C over the Collinder~228 (South Pillars) area.  Region D represents all other widely distributed stars in the full sample; this group has $> 5000$ probable Carina members.  All 14,368 CCCP sources lie in one of these four Regions; for example, Region A includes members of Trumpler~14 and 15 as well as members in the surrounding area.  The observed group member tallies and average absorptions of these regions are listed at the end of Table 1.

\section{Cluster Memberships \label{clusmem.sec}}

To document the clustering results reported here, and to assist further study of individual clusters and groups, Table 3 lists the cluster, group, and region membership for all 14,368 sources in the CCCP catalog.  Columns (1)$-$(8) reproduce useful source properties from \citet[][Table~1]{Broos11a}: CCCP sequence number, label and IAU-designated name; right ascension and declination; net extracted counts; the $F_{t,photon}$ nonparametric measure of photon flux in the $0.5-8$~keV band (\S~\ref{methods.sec}); and background-corrected median energy. Column (9) reproduces from \citet{Broos11b} the likely classification of each source, based on X-ray, infrared, and positional properties: class H1 is a probable foreground star; H2 is a probable Carina member; H3 is a probable background star; H4 is a probable extragalactic contaminant; and H0 is an unclassified source.  Columns (10) and (11) report the cluster memberships obtained here. Designations ``C1'', ``C2'', $\ldots$, ``C20'' refer to the clusters in Table~1.  Designations ``G1'', ``G2'', $\ldots$, ``G31'' refer to the groups in Table~2.  Designations A, B, C, and D refer to the large-scale regions discussed in \S~\ref{regions.sec}. 

Recall that the samples studied here were restricted to the H2-class sources (\S~2) and the classification procedure is unreliable in many cases.  Particularly in the large-scale Regions A-D, it is quite possible that some CCCP sources classified as probable Carina members (class H2) are in fact foreground (class H1) or background (class H3) Galactic field stars.  Validation of the classifier using HAWK-I color-magnitude diagrams \citep{Preibisch11} suggests that we probably overestimate the population of Carina members and underestimated the population of field star contaminants by a few percent \citep[][\S~5]{Broos11b}.   Conversely, some individual H1 and H3 stars are likely true Carina members, as are many of the unclassified sources (class H0).  We thus caution that the individual source classifications given in Table 3 represent our best knowledge, but will have a considerable number of errors.

\section{Results and Discussion} \label{disc.sec}

The map of spatially complete probable Carina members from the CCCP survey in Figure~\ref{Complete_smooth.fig} and the associated list of clusters in Table~1 are the principal empirical results of this study.  They give a much clearer view of the stellar concentrations than available from historical optical studies which are confused by complicated distributions of dust absorption and nebular emission.

\subsection{The historic Trumpler clusters}

The three principal clusters --- Trumpler~14, 15, and 16 --- are distinct from each other, with strong peaks in stellar surface density separated by substantial dips.  The \Chandra\ data do not support the suggestion that Trumpler~14 and 16 are contiguous, separated only by a dust lane \citep{Feinstein73}.   A broad bridge of stars is seen between Trumpler~14 (CCCP-Cl~1) and Trumpler~15 (CCCP-Cl~8), which is part of the Region~A large-scale density enhancement.  This bridge strongly indicates that Trumpler~15 is part of the Carina complex rather than the chance superposition of a more distant cluster as suggested by several authors \citep[e.g.][]{Walborn95}.   CCCP results on Trumpler~15 and its relation to the Carina complex are discussed in detail by \citet{Wang11}. 

Each of these three rich clusters have internal structure which cannot be readily summarized by simple terms such as a circular or elliptical shape.  In both Trumpler~14 and 15, the peak in stellar surface density is offset from the center of outer density contours by $0.5-1$~pc, and the distribution of lower mass members is not symmetrically distributed around the dominant O stars.  The internal structure of Trumpler~15 is discussed by \citet{Wang11}.  The stellar structures may be related to nearby molecular cloud cores \#10 and \#11 studied by \citet{Yonekura05}.  Spherically symmetric models (for example, as discussed by \citet{Ascenso07} for Trumpler~14) will not capture the full complexity of the star distribution.  

The internal structure of Trumpler~16 is qualitatively different from Trumpler~14 and 15, as it does not have a single surface density peak.  Although our analysis divides it into seven clusters (CCCP-Cl~3, 4, 6, 9, 10, 11, and 12), this number of subclusters identified depends strongly on the smoothing kernel.  The structure seen in  Figure~\ref{Complete_smooth.fig} can be described as hierarchically clustered.  For example, one could choose to define a single cluster centered at ($10^h44^m50^s, -59^\circ43\arcmin$) with diameter 11\arcmin\/ (7.3~pc) containing several subclusters, none of which lies at the cluster `center'.   The curved southern boundary of the cluster may, in part, be caused by absorption in the {\bf V}-shaped dust lane that crosses the middle of the Carina field.  See Figure 1 of \citet{Albacete09} for the spatial relationship between the dust lane and the X-ray population.   \citet{Evans03} and \citet{Albacete09} give detailed \Chandra\ studies of the X-ray properties of Trumpler~16 stars.  CCCP results on Trumpler~16 are discussed in detail by \citet{Wolk11}.

\subsection{The historic Collinder and Bochum clusters} \label{Collinder.sec}

The Figure~\ref{Complete_smooth.fig} map clarifies the long-standing confusion regarding the status of less rich clusters in the Carina complex (\S 1).  We find that Collinder~232 east of Trumpler~14 is a distinct, compact cluster (CCCP-Cl~5), confirming the conclusion of  \citet{Carraro04}.  It is clear that Collinder~232 is not just part of the halo of Trumpler~14.  Note that, as in the richer Trumpler clusters, its lower mass stars are not centered on its dominant O stars.  Infrared studies show that Collinder~232 stars have a wide range of absorptions and a considerable population of very young protostars \citep{Preibisch11, Povich11b}.  

Several researchers have commented that Collinder~228 in the South Pillars region is not a distinct cluster, but may be an extension of Trumpler~16 \citep[e.g.][]{Walborn95, Smith08}.  The X-ray source distribution in Figure~\ref{Complete_smooth.fig} does not support this interpretation, as the stellar surface density of Trumpler~16 shows a distinct near-circular boundary with diameter around 11\arcmin.  

Instead, the X-ray source distribution in the southern part of the Carina complex supports the idea that Collinder~228 is a loose association of young stars and sparse groups distributed over a very large ($> 20\arcmin$\/ diameter)  region centered around QZ Car at ($10^h44.4^m -60^\circ00\arcmin$), as described by \citet{Feinstein76}.  Collinder~228 might be considered to be the stars associated with the dusty South Pillars cloud structures \citep{Smith00, Smith10}.  Our Region~C of enhanced X-ray star densith would then trace the lower mass stars in the inner $10\arcmin$\/ of the cluster\footnote{Neither the X-ray sources nor the historical optical photometric samples show a concentration of young stars at the nominal location of Collinder~228 listed in open cluster catalogs: ($10^h43.8^m, -60^\circ05\arcmin$) \citep[][and the SIMBAD database]{Kharchenko05}, ($10^h42.1^m,-59^\circ55\arcmin$) (the WEBDA database on open clusters, \url{http://www.univie.ac.at/webda}), or ($10^h43.2^m,-60^\circ00\arcmin$) \citep[the 2MASS location given by][]{Dutra03}. These locations are several arcminutes west of our stellar groups in Region C, and we do not see any enhancement in young star densities at these locations.   We suggest that the optical and near-infrared surveys are not detecting a physical young stellar cluster at their stated locations, but rather are concentrating on local groups of massive stars within the more extended distribution described by \citet{Feinstein76} and seen as Region C of the X-ray selected population.}.  Examination of Figure~\ref{Full_maps.fig} shows the low mass star distribution attributable to Collinder~228 is not uniform but exhibits distinct clumping.  The most prominent grouping is around ($10^h45.6^m, -59^\circ58\arcmin$) which includes CCCP-Cl~13, 17 (the Treasure Chest), 18, and CCCP-Gp 21, 22, 23, 24, 25, 26, and 27.  Another concentration is a $\sim 8$\arcmin\/ diameter region around ($10^h44.0^m,-60^\circ00\arcmin$) where we find CCCP-Gp 8, 9, 10, 11, 12, 13, 14, 16, and CCCP-Cl~2. Group~14 is centered on the well-studied multiple supergiant system QZ~Car (O9.7 I + O8 III, V=6.3) studied by \citet{Parkin11}.  Other groups are distributed up to $\sim 20\arcmin$\/ south of the Treasure Chest including CCCP-Cl~16, 19, 15, and CCCP-Gp 18.  These stellar groupings are also independently found in the {\em Spitzer} mid-infrared study of \citet{Smith10}.  We caution that the Collinder~228 region considered here covers several ACIS fields, so the X-ray sensitivity is not uniform.  In particular, sparse groups of lower mass stars that happen to fall far off-axis in the ACIS observations may be missed.  

Collinder~234 is a poorly studied cluster lying $\sim 3$\arcmin\/ southeast of CCCP-Cl~12, the highest surface density spot of X-ray stars in the Trumpler~16 complex.  It is reportedly as a concentration of $\ga 10$ stars with diameter 4\arcmin\/ in the 2MASS survey centered at ($10^h45.2^m, -59^\circ45\arcmin$) \citep[][]{Dutra03}.   A cluster at the Collinder~234 location is not confirmed in the X-ray sample of PMS stars.  The CCCP sample does include several optically bright late-O stars \citep[][and SIMBAD]{Gagne11}: LS~1870 (V=10.0, O9 III);  MJ \#484 (V=10.0, O9 III);  MJ \#496 (V=11.0, O8.5 V); HD~93343 (V=9.6, O8 V + O 7-8.5 V);  MJ \#516 (V=9.3, O8 V + O9.5 V); MJ \#517 (V=9.3, O7 V + O8 V + O9 V); and MJ \#535 (V=9.3, O5.5-O6 V(n)((fc)) + B2 V-III).  Except for a sparse subcluster of a few lower mass stars within a few arcseconds of MJ~\#496, there are no X-ray emitting PMS stars in the vicinity of these massive stars.   This is quite extraordinarily different from other young clusters.  The Orion Nebula Cluster has a similar distribution of late-O stars and would have $\sim 360$ PMS stars detected in the CCCP were it to lie in the Carina Nebula \citep{Townsley11a}.

Bochum~11 to the southeast is confirmed in the X-ray cluster list (CCCP-Cl~20) as a sparse cluster with low stellar surface density.  Examination of Figure~\ref{Full_maps.fig} suggests that it consists of two star groupings in a $\sim 6\arcmin$\/ (4~pc) diameter region: one containing CCCP-Cl~20 and CCCP-Gp 31 around ($10^h14^m16^s,-60^\circ07\arcmin$); and the other containing CCCP-Gp 28, 29, and 30 around ($10^h46^m52^s, -60^\circ05\arcmin$).   Group~28 lies $\sim$30\arcsec\/ northwest of the spectroscopic binary HD~93576 (V=9.7, O9 IV) and appears  associated with the mid-infrared source MSX6C~G287.9816-00.8724. Group~30 is centered on HD 93576 and appears associated with MSX6C~G287.9475-00.8846.

\subsection{Obscured and embedded clusters}

Highly luminous and extended mid-infrared sources, such as $IRAS$ sources with bolometric luminosities in the $L \sim 10^4-10^6$~L$_\odot$ range, and compact radio H~II regions are relatively rare in the Carina complex compared to other active high-mass star forming regions.  This demonstrates that few embedded clusters with luminous OB stars.   However, two infrared sources associated with current star formation appear as X-ray selected clusters.  

First, CCCP-Cl~17 is the Treasure Chest, a compact cluster with bright mid-infrared emission dominated by an O9.5 star producing the compact HII region MSX~G287.84-0.82 \citep{Hagele04, Smith05}.   About 100 members are in the full X-ray sample.  Their average X-ray median energy is 1.8~keV, only slightly higher than the rich lightly obscured clusters; this is surprising considering the $A_V \sim 10-50$ found from infrared studies of some Treasure Chest members.    This suggests that Treasure Chest stars have a wide range of absorptions.  

Second, a sparse grouping of CCCP sources can be seen on the eastern side of the Treasure Chest cluster near the compact mid-infrared source IRAS~10439-5941 = MSX6C~G287.84-0.82, a very bright illuminated pillar in mid-infrared images of the Carina Nebula \citep{Smith00, Rathborne04}.  It is a faint radio continuum compact HII region and is associated with a molecular globule with column density $N_{H_2} \simeq 9 \times 10^{21}$ cm$^{-2}$ and mass $30-200$ M$_\odot$. The infrared source has a very high bolometric luminosity $3.7 \times 10^6$ L$_\odot$ and likely represents the collective emission of an embedded OB cluster.  The CCCP survey has detected a few of these stars.   

Third, Group~20 just south of Trumpler~16 is centered 30\arcsec\/ west of IRAS~10430-5931, an embedded protostar with bolometric luminosity $\sim 1 \times 10^4$ L$_\odot$ accompanied by a compact group of lower-mass stars near the edge of the bright-rimmed molecular globule MSX6C~G287.63-0.72.  This small star forming region is discussed in detail by \citet{Megeath96}.  The strongest X-ray source here is a high-$L_x$ PMS star presented by \citet{Evans03}.  

Other sparse groups of X-ray sources that did not satisfy our criteria for producing the lists in Tables~1 and 2 may be present.  For example, a few sources appear to be associated with the mid-infrared source IRAS~10441-5949 = MSX6C~G287.8893-00.9316, which is modeled as a dusty proto B0 star with bolometric luminosity $\simeq 3 \times 10^4$ L$_\odot$ by \citet{Rathborne04}.   

The inferred fraction of embedded X-ray stars is relatively low compared to other large-scale star formation regions studied with \Chandra.  For example, the M~17, Rosette, and NGC~6634 complexes each have hundreds of embedded X-ray stars, a significant fraction of the population in the associated revealed central clusters \citep{Broos07, Wang09, Feigelson09}.  The comparison with the Rosette Molecular Cloud is particularly valuable, as the CCCP ACIS exposures have sensitivities similar to the RMC maps presented by Wang et al.\ where a number of obscured clusters, each with dozens of X-ray stars, were readily detected. Were a similar ratio of obscured-to-revealed stars present in the Carina complex, thousands of heavily absorbed stars would have been detected with cluster/group $\langle MedE \rangle$ values in the range $2-4$~keV.   Nonetheless, a moderate population of nascent stars is present:  $\sim 1400$ infrared-excess young stellar objects is found from {\it Spitzer} IRAC maps, most of which are not detected in X-rays \citep{Povich11b}.

\subsection{New star groups}

\begin{enumerate}

\item CCCP-Cl~14 lies southeast of Trumpler~16, centered on the eclipsing binary O5.5V star V662~Car (Figure~\ref{Full_maps.fig}c).  \citet{Sanchawala07} first reported an X-ray selected cluster here, identifying 10 of the 41 CCCP sources we assign to the cluster.  Lying 2\arcmin\/  northeast of the C$^{18}$O cloud core \#12, its CCCP members have a somewhat higher absorption ($A_V \simeq 10$ mag; Table 1) than that reported by \citet{Sanchawala07} for the other principal clusters in the Carina complex. 

\item CCCP-Gp 18, lying at the southern tip of the mosaic, has only five faint X-ray source members, but is extraordinary in two ways \citep{Townsley11a}.  Three of the sources lie within $\sim 1$\arcsec\/ of each other, and the group's median energy, $\langle MedE\rangle > 5$~keV, is equivalent to $A_V \sim 125$ mag.  \Chandra\/ observations of star forming regions have, on rare occasions, detected sources with median energies around $5-6$~keV. Examples include several stars near the Becklin-Neugebauer hot core in Orion \citep{Grosso05}, several stars in  NGC~6334~I(N) and nearby embedded clusters \citep{Feigelson09}, and two Class 0/I protostars in nearby star forming cores \citep{Hamaguchi05, Getman07}.  However, there have been no reports of a molecular cloud, diffuse or compact infrared emission, or other indicator of current star formation at the location of Group~18 ($10^h44^m51.9^s, -60^\circ25\arcmin09\arcsec$). 

\item CCCP-Gp 4 and 7 can be seen as low contours $\sim 10$\arcmin\/ northwest of Trumpler~14 in Figure~\ref{Complete_smooth.fig}.  Group~4 is associated with the faint mid-infrared source MSX6C~G287.2238-00.5339 and the C$^{18}$O cloud core \#9 with estimated molecular gas mass $1400-2200$ M$_\odot$ and column density $\log N_{H_2} \simeq 21.9$ cm$^{-2}$ \citep{Yonekura05}.  Three prominent 2MASS stars are also present, two of which are young stars with infrared-bright disks \citep{Mottram07}.   Group~4 members have a high $\langle MedE\rangle$ equivalent to $A_V \sim 11$, consistent with being embedded in this cloud.  These groups are discussed in \citet{Povich11b}. 

\end{enumerate}

\subsection{Large-scale and Unclustered Star Distributions}

The origin of the dispersed distribution of young stars discussed in \S~\ref{regions.sec} is not obvious, and several causes may be present.  The stars in Regions A and B may be associated with the Trumpler~14, 15, and 16 clusters, either as primordial low density halos or as members ejected by dynamical interactions.  Evidence for a stellar halo around Trumpler~14 also emerged from the near-infrared study of the inner 3\arcmin\/ by \citet{Ascenso07}.  The star distribution in Region C was discussed in \S~\ref{Collinder.sec}.  A few groups are associated with current star formation in the Southern Pillars, and we suspect that other members were formed in similar molecular structures that have now dissipated.  Further discussion of this possibility appears in a recent {\it Spitzer} study of stars in the Southern Pillars \citep{Smith10} and in {\it Chandra} studies of star formation around nearby HII regions \citep[e.g.][]{Getman07}.   

The widely distributed stars (Region D) do not exhibit any concentrations or association with prominent young clusters.  Their distribution is likely to extend beyond the $\sim 50$~pc CCCP mosaic.  These most likely represent an older population of pre-main sequence stars, much as the Orion Nebula region (Ori OB1d) is surrounded by older, more dispersed clusters \citep[Orion OB 1a-c in][]{Bally08}.  A widely distributed population of massive stars was also found in a \Chandra\/ study of the NGC~6334 complex \citep{Feigelson09}.  The idea of a previous generation of star formation whose most massive members have produced supernovae in Carina is strongly indicated by the strong diffuse X-ray emission \citep{Townsley11b}, the presence of a young neutron star in Region D east of Trumpler 16 \citep{Hamaguchi09}, and the absence of early-O stars in Tr~15 \citep{Wang11}.

We can quantify the size of the populations in the four large-scale Regions by constructing X-ray luminosity functions (XLFx) of CCCP sources classified as Carina members.  These are shown in Figure~\ref{regions_xlf.fig} for each Region, where a small number of highly luminous sources associated with known OB stars \citep{Skiff09} are removed to facilitate comparisons among lower-mass populations.  Approximate observed X-ray luminosities for individual sources are obtained by multiplying the photon flux $F_{t,photon}$ (\S~\ref{methods.sec}) times the median energy and the factor $4 \pi d^2$ where $d=2.3$~kpc.   We then apply an absorption correction based on the median energy value, as described by \citet{Getman10}.  This correction is not large in most cases, but is not available for the faintest sources.   The Carina Region XLFs are compared to the XLF of 839 lightly-obscured low-mass members of the Orion Nebula Cluster (ONC) obtained in the {\it Chandra Orion Ultradeep Project} \citep[COUP,][]{Getman05, Feigelson05}.  The membership and Initial Mass Function of the ONC is accurately measured \citep[e.g.][]{Muench02}, so it can serve as a calibrator for more distant clusters.  We assume the ONC population above the brown dwarf limit ($M > 0.08$~M$_\odot$) is 2,800 stars \citep{Hillenbrand98}.  

The CCCP XLFs have a similar shape to the COUP XLF above $\log L_x \simeq 30.5$ erg~s$^{-1}$;  the XLFs in this luminosity regime are roughly a powerlaw $dN/d(\log L_x) \propto -1.0 ~{\rm to}~ -0.9) \times \log L_x$.  We then examine the vertical offsets between the fitted powerlaws in Figure~\ref{regions_xlf.fig} to estimate the total populations of young stars in each region.   The membership and Initial Mass Function of the ONC is measured \citep[e.g.][]{Muench02}, so it can serve as a calibrator for more distant clusters.  We assume that the ONC population is 2800 stars, obtained by Hillenbrand \& Hartmann (1998) within a radius of 2.06~pc.  The total population, including brown dwarfs and outer regions of the cluster, may be substantially larger.

The CCCP XLFs have a similar shape to the COUP XLF above $\log L_x \simeq 30.5$ erg~s$^{-1}$;  the XLFs in this luminosity regime are roughly a powerlaw $dN/d(\log L_x) \propto -1.0 ~{\rm to}~ -0.9) \times \log L_x$.  We then examine the vertical offsets between the fitted powerlaws in Figure~\ref{regions_xlf.fig} to estimate the total populations of young stars in each region.  The results are Region A has 10 ONCs (28,000 stars); Region B has 6 ONCs (17,000 stars); Region C has 5 ONCs (14,000 stars); and the outlying areas in Region D have 16 ONCs (45,000 stars). The total population in the CCCP field is then about 104,000 stars down to the brown dwarf limit.

Caution is needed in interpreting these estimates of total Carina member populations.  The XLFs will be significantly affected by the strong spatial variations of the survey sensitivity on these scales which combine several ACIS pointings.  We have further assumed that the classification of individual CCCP sources into Carina members and contaminants has no error, and that none of the `unclassified' sources are members.  Both of these assumptions are not accurate \citep{Broos11b}.  CCCP sources with very few photons do not have estimated X-ray luminosity values and are missing from the lower end of the  XLFs.  The 104,000 star estimate undercounts heavily obscured populations that are not well-represented in the CCCP survey \citep{Povich11b}.  Nonetheless, because these estimates of total young stellar population are scaled to the ONC where the membership is well-established, they represent the estimates of the full populations down to the brown dwarf limit, not just the X-ray detected sample.

\section{Conclusions}

The \Chandra\/ Carina Complex Project provides a new window on the clustering structures in the Carina Nebula, arguably the nearest laboratory for studying starburst phenomena seen in other galaxies.  The X-ray data, particularly the smoothed map of the spatially complete sample in Figure~\ref{Complete_smooth.fig}, elucidate clustering structure that is difficult to visualize from optical (where patchy dust absorption and nebular emission dominate) and infrared (where dust emission and field star contamination dominate) images.  While the detailed lists of clusters, groups, and stellar memberships are not uniquely derived, the overall picture of the Carina `cluster of clusters' emerges. 

The main result of astrophysical interest is that different types of clusters appear to be present.  First, Trumpler~14 and 15 might be called `classical' young stellar clusters which are likely to evolve into OB associations.  They have rich stellar populations in a unimodal, centrally concentrated structure several parsecs across.  Further study is needed to understand the non-spherical distributions and off-center massive stars.  Trumpler~16 represents a second class of rich cluster with a multimodal structure. As the stellar crossing-time is probably several million years, its clumpy structure may represent an unequilibrated stage of cluster formation.   Collinder~228 is a third type of cluster which extends over tens of parsecs with many sparse compact groups without any rich concentration.   We suspect that these groups arose from triggered star formation processes over an extended time period in illuminated cloud structures like the South Pillars as discussed in more detail by \citet{Smith10, Povich11b}.

The large-scale, low surface density distributions in Regions A and B may be a fourth type of clustering structure.  The O stars historically attributed to Collinder~234 that do not have an associated concentrated population of PMS stars may be part of this dispersed population.  It is remarkable that the dispersed stars in Region A ($\sim 2900$ detected stars) are more numerous than those inside our boundaries delineating Trumpler~14 and Trumpler~15 ($\sim 1900$ detected stars). The origin of these large-scale stellar structures is not obvious.  The widely dispersed population far from all young clusters (Region D) is estimated to have 45,000 young stars.   Some may have been dynamically ejected from dense cores of concentrated rich clusters, and others may be evaporated from sparse groups.  They may have formed in a dispersed triggered fashion similar to the Collinder~228 stars in the South Pillars today, in which case the triggered population may exceed the clustered population.  Or they may represent an early generation of star formation from cloud material that is now dispersed, analogous to the older Orion clusters around the Orion Nebula Cluster.   Observational constraints on the ages of the dispersed population would be very helpful in discriminating their origin.      

The X-ray luminosity functions of the large-scale regions of the Carina complex are estimated and scaled to the Orion Nebula Cluster.  Although only approximate, we estimate that the entire CCCP field has $\sim$104,000 young stars of which only half lie in concentrated clusters.

A final important result is that most CCCP X-ray sources exhibit only light absorption, much of which is due to foreground line-of-sight material.  X-ray clusters and groups are not concentrated in the molecular clouds mapped in CO line emission \citep{Yonekura05}, and the CCCP finds fewer embedded stars and clusters than are reported in $Chandra$ studies of other giant molecular clouds.  As a considerable population of infrared-excess protostars are found by \citet{Povich11b}, this suggests that the CCCP is not sufficiently sensitive to detect most of the younger, embedded protostars. 
 
The CCCP thus reveals considerable complexity in clustering structures and processes.  These can be studied in more detail within Carina, and insights from Carina can be extended to further our understanding of starburst phenomena in broader contexts.

\acknowledgments  Acknowledgements.  We thank Cathie Clarke (Cambridge) and Nicolas Grosso (Strasbourg) for discussions, and the anonymous referee for useful suggestions.  This work is supported by Chandra X-ray Observatory grant GO8-9131X (PI: L. Townsley) and by the ACIS Instrument Team contract SV4-74018 (PI: G. Garmire), issued by the Chandra X-ray Center, which is operated by the Smithsonian Astrophysical Observatory for and on behalf of NASA under contract NAS8-03060.  K.G.S. is supported  by NSF grant AST-0849736.

\facility{CXO (ACIS)}

\begin{figure}
\centering
\includegraphics[width=1.0\textwidth]{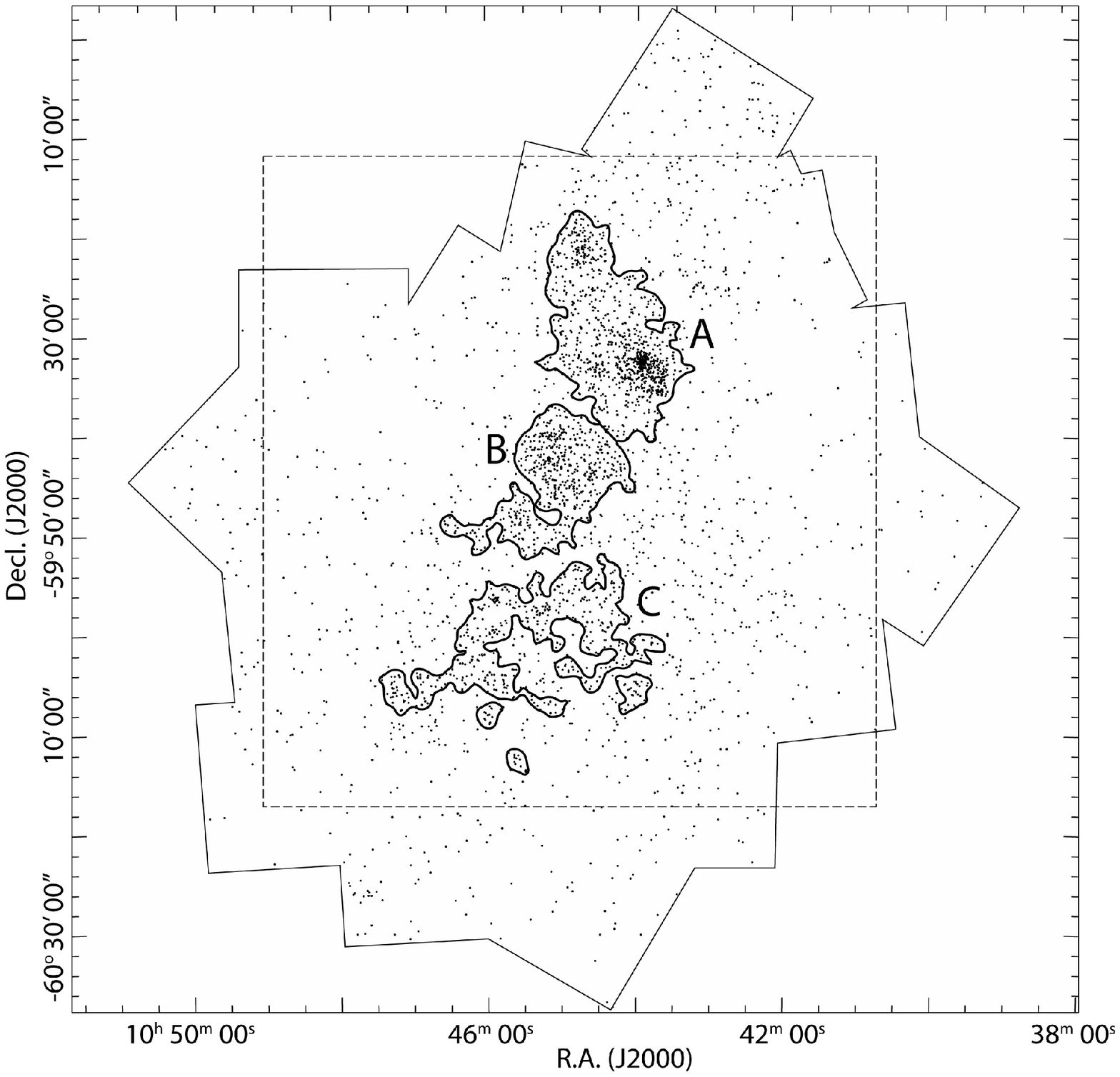} \\  
\caption{Spatially complete sample of 3,220 probable Carina members selected from the CCCP X-ray survey.  (a) X-ray source distribution.  The CCCP field of view is outlined by the polygon.   The boundaries of the large-scale stellar enhancements A, B, C (\S 3.3) are marked by the contours. The dashed rectangle demarcates the central $62\arcmin \times 65\arcmin$ field shown in panel $b$. (b) Contour map of the X-ray source surface density smoothed with a $\sigma = 30$\arcsec\/ Gaussian.  Twenty clusters above the third contour and three large-scale enhancements are marked.  The contours are in linear units of surface density from 1 to 8 units of surface density. }  \label{Complete_smooth.fig}t
\end{figure}

\begin{figure}
\centering
\includegraphics[width=1.0\textwidth]{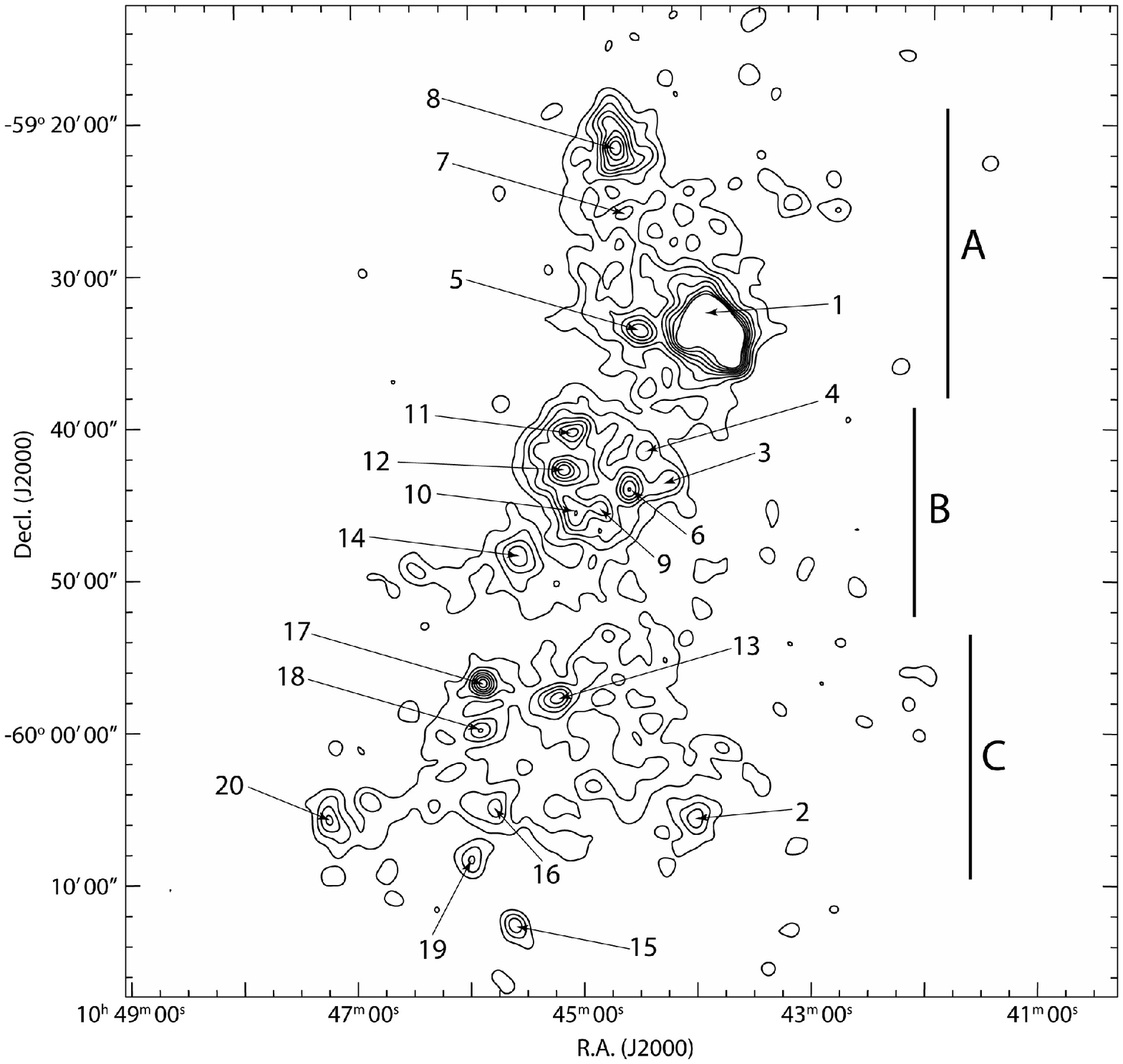}
\end{figure}

\begin{figure}
\centering \vspace*{-0.5in}
\includegraphics[width=0.8\textwidth]{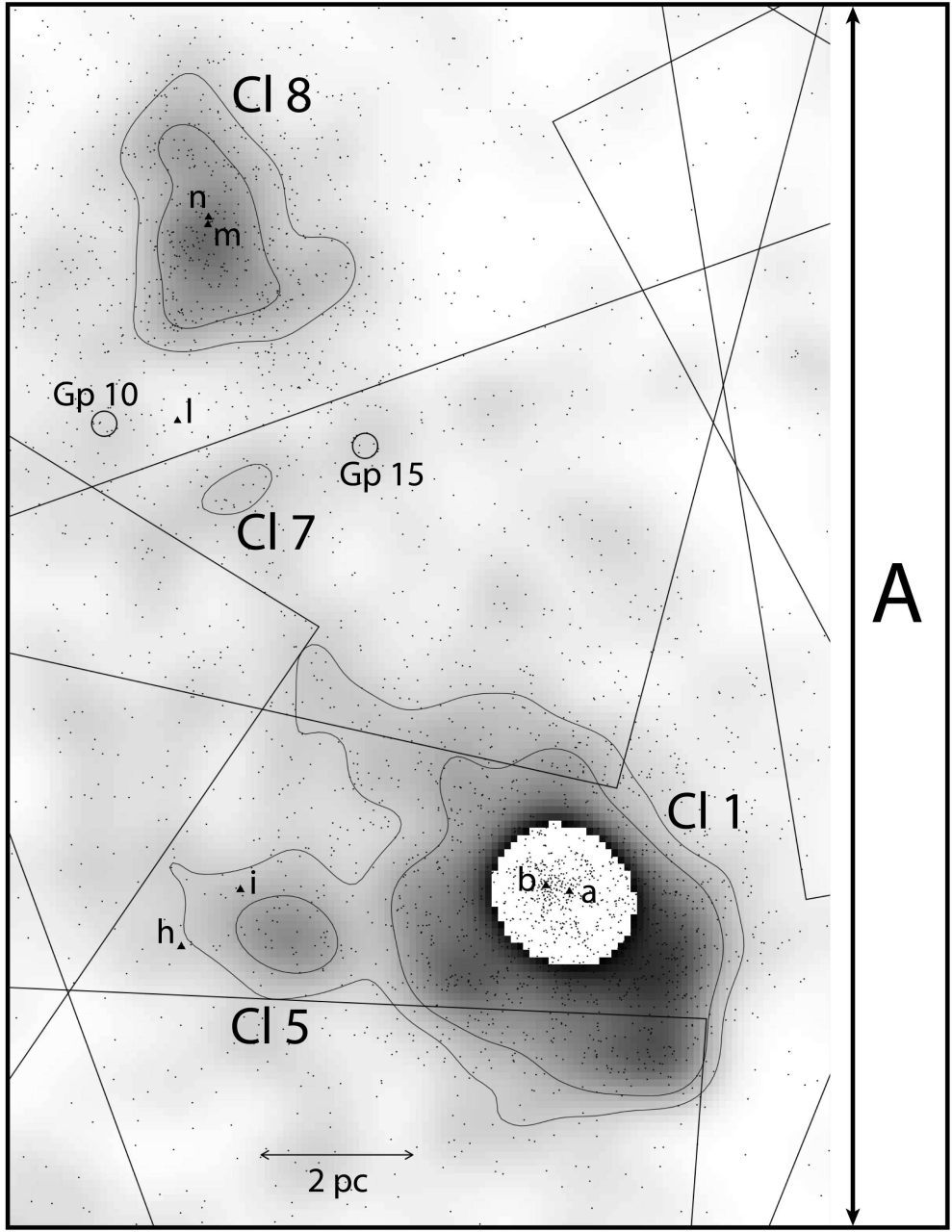} \\ 
\caption{Stellar spatial distribution in three clustered regions of the Carina complex using the full sample of 10,728 X-ray Carina members (small dots).   Trumpler~14 and Trumpler~15 region (top left panel); Trumpler~16 region (top right panel); and  the South Pillars region (bottom panel).  The gray-scale map with two contours (0.003 and 0.005) are reproduced from Figure 1b using the spatially complete sample. Principal clusters from Table 1 are labeled Cl 1-20, and small groups from Table 2 are located with circles and labeled Gp 1-31. Dominant stars in major clusters are shown as filled triangles and labeled $a-y$ as indicated in Table 1.  Overlapping rotated squares outline the {\it Chandra} ACIS fields. The extents of the large-scale stellar enhancements in declination are indicated on the right sides, and the 2~pc scale bar assumes a distance of 2.3~kpc to the complex. } \label{Full_maps.fig}
\end{figure}
\newpage

\begin{figure}
\centering
\includegraphics[width=0.7\textwidth]{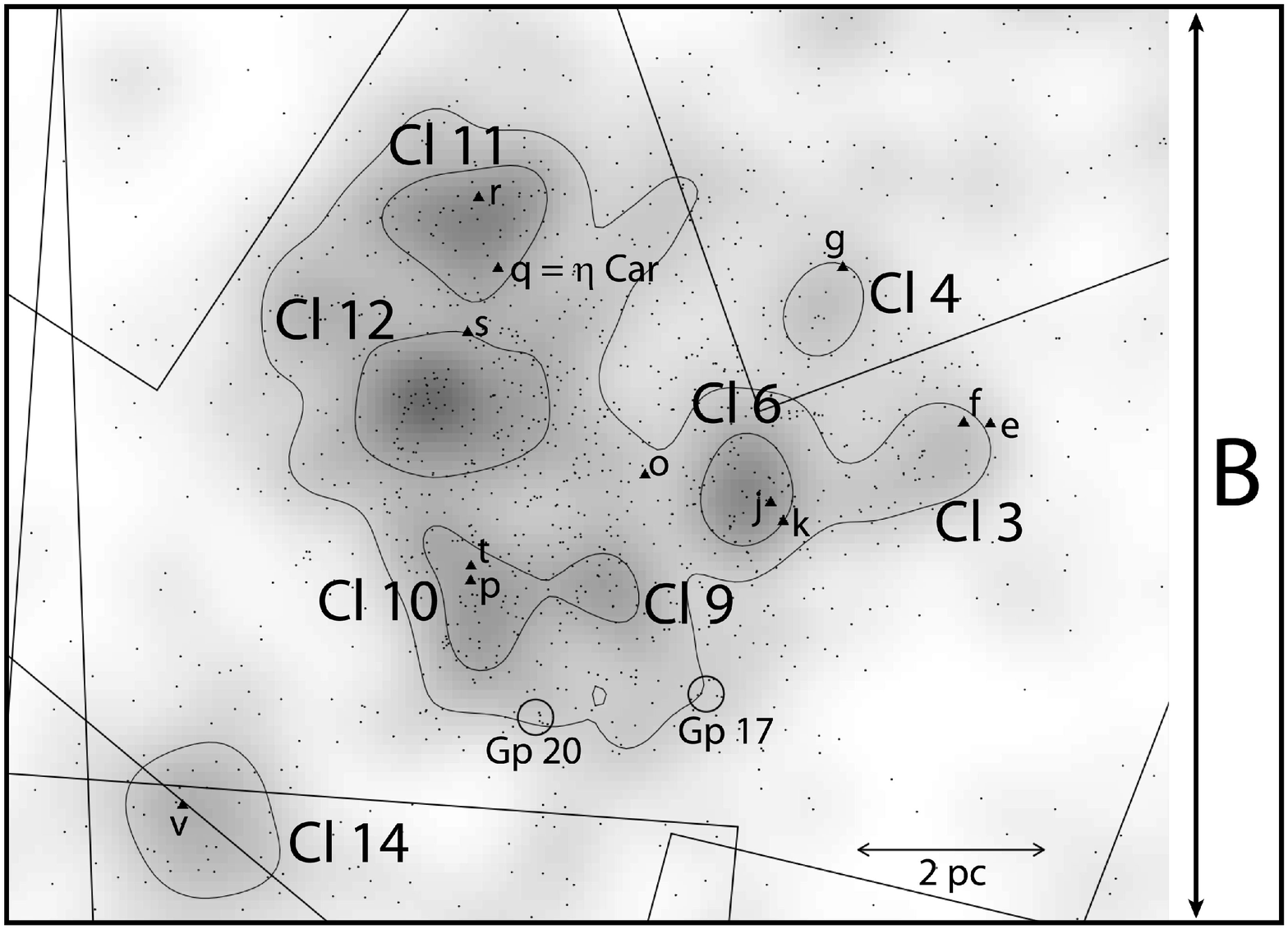} 
\end{figure}

\begin{figure}
\centering
\includegraphics[width=0.9\textwidth]{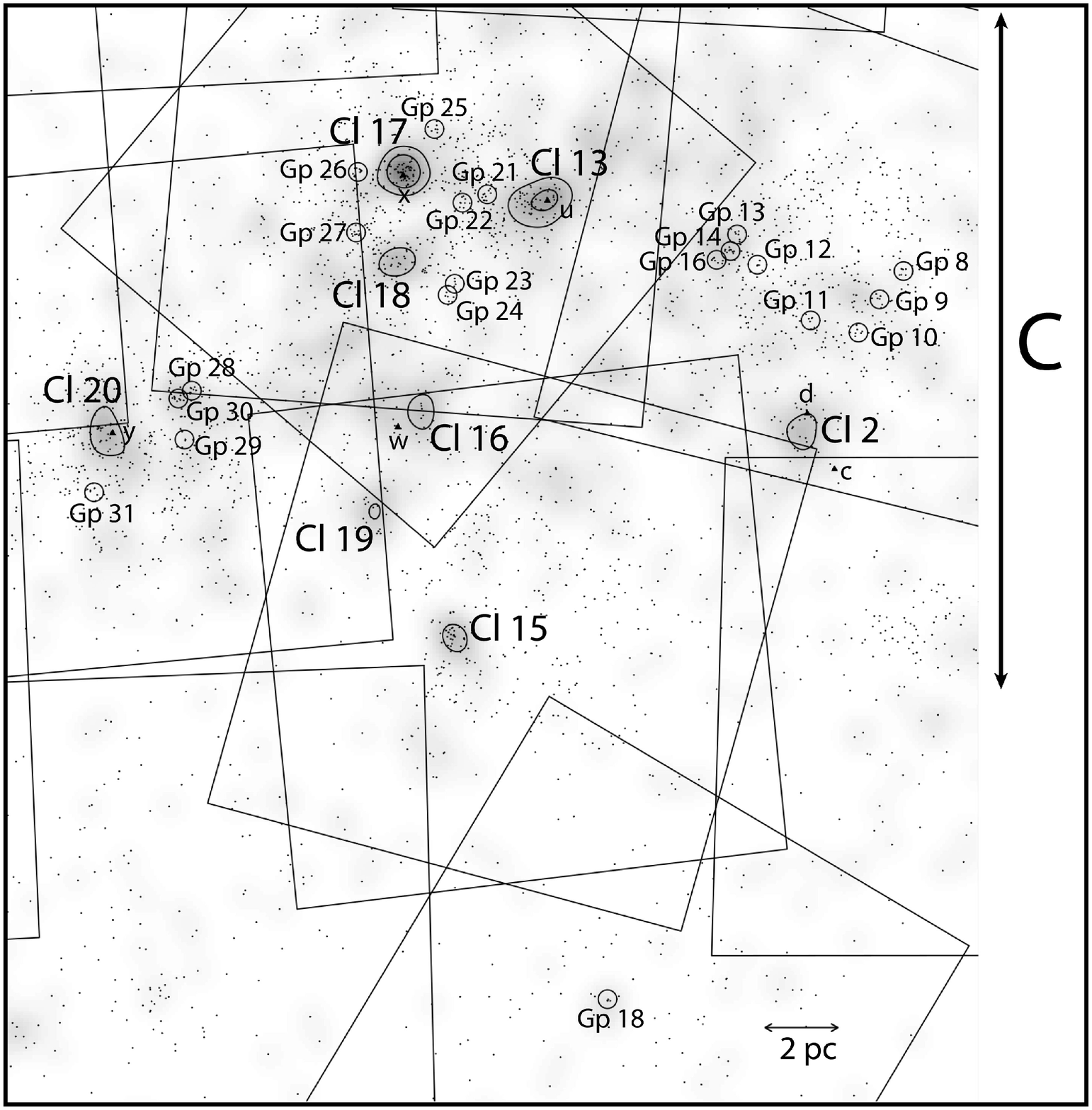} 
\end{figure}

\begin{figure}
\centering
\includegraphics[width=1.0\textwidth]{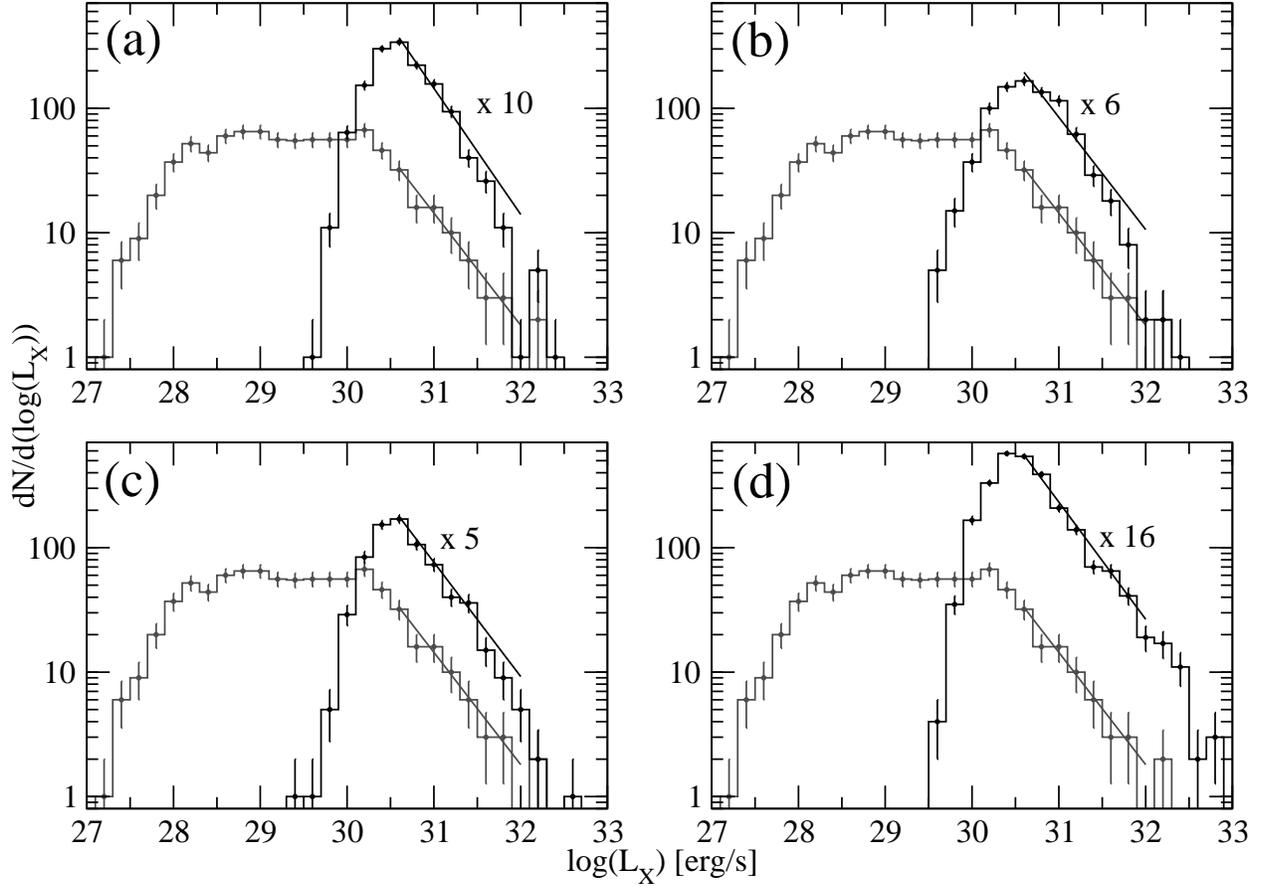} 
\caption{Estimated X-ray luminosity functions (upper histograms) for the CCCP sources classified as likely Carina members (excluding OB stars) in the large-scale (a) Region A, (b) Region B, (c) Region C, and (d) Region D.  The lower histograms show the XLF of the low-mass COUP Orion Nebula Cluster, and lines show powerlaw fits.} \label{regions_xlf.fig}
\end{figure}

\begin{deluxetable}{crrllrrlrlll} 
\tablecaption{Principal clusters of X-ray stars in the Carina star forming complex (\S~\ref{majclus.sec})
}
\tabletypesize{\scriptsize}
\rotate
\centering
\tablewidth{0pt}
\tablecolumns{12}
\tablehead{
\multicolumn{5}{c}{X-ray selected cluster} && \multicolumn{3}{c}{absorption} && \multicolumn{2}{c}{Optical counterpart} \\
\cline{1-5} \cline{7-9} \cline{11-12}
\colhead{CCCP-Cl} & \multicolumn{2}{c}{RA ~(J2000)~~Dec} &&  
\colhead{Cntr} & \colhead{$N$} & \colhead{$N_{ME}$} & \colhead{$\langle MedE\rangle~(sd)$} & \colhead{$A_V$} && 
\colhead{Opt/IR cluster} & \colhead{Dominant stars}  \\
& & && & & & keV & mag && & \\ 
\colhead{(1)} & \colhead{(2)} & \colhead{(3)} & \colhead{(4)} & \colhead{(5)} && \colhead{(6)} & \colhead{(7)} & \colhead{(8)} && \colhead{(9)} &\colhead{(10)}  
}
\startdata
~1 & 10:43:57.0 & -59:32:52&& 2m& 1378 & 1138 & ~~1.56~~( 0.48) &  3~~ && Trumpler~14 & HD 93128 (O3.5 V((fc)),a), HD 93129A (O2 If*,b)   \\
~2 & 10:44:01.1 & -60:05:46  && 1   &     48 &     48 & ~~1.43~~( 0.53) &  2~~ && {\em Spitzer} A $\star$ &  WR 24 (WN6,c), HD 93146 (O7 V((f)),d)  \\
~3 & 10:44:16.0 & -59:43:37 && 3m &     32 &     32 & ~~1.55~~( 0.43) &  3~~ && Trumpler~16   & HD 93162 (WN,e), Tr16-244 (O3/4 If)\\
~4 & 10:44:28.5 & -59 41:38 && 3    &     10 &     10 & ~~1.47~~( 0.12) &  2~~ && Trumpler~16   & CPD-59$^\circ$2574 (B IVn,g)\\
~5 & 10:44:31.3 & -59:33:44 && 3m &     70 &     70 & ~~1.55~~( 0.50) &  3~~ && Collinder~232 & HD 93250 (O4 III(fc),h), HD 303311 (O5 V,i) \\
~6 & 10:44:36.5 & -59:44:10 && 3m &   109 &     98 & ~~1.39~~( 0.34) &  2~~ && Trumpler~16   & HD 93205 (O3 V + O8 V,j), HD 93204 (O5.5 V((fc)),k) \\
~7 & 10:44:39.2 & -59:26:02 && 2m &     44 &     41 & ~~1.43~~( 0.27) &  2~~ && \nodata & RT Car (M2 Iab,l) \\
~8 & 10:44:43.8 & -59:21:42 && 2    &   481 &   331 & ~~1.43~~( 0.52) &  2~~ && Trumpler~15   & HD 93249 (O9 III,m), CD-58$\circ$3536B (O9 III,n)\\
~9 & 10:44:51.6 & -59:45:26 && 4m &     55 &     47 & ~~1.73~~( 0.71) &  5~~ && Trumpler~16    & Tr16-104 (O7V((f)) + O9.5 + B0.2 IV,o), Tr16-23 (O9 III,p) \\
10 & 10:45:04.6 & -59:45:44 && 4m &     84 &     84 & ~~1.59~~( 0.53) &  3~~ && Trumpler~16   & Tr16-23 (O9 III,p) \\
11 & 10:45:06.1 & -59:40:27 && 4m &     71 &     71 & ~~1.58~~( 0.44) &  3~~ && Trumpler~16   & $\eta$ Car (LBV,q), HD 303308 (O4 V((fc)),r) \\
12 & 10:45:10.4 & -59:42:55 && 4m &   169 &   159 & ~~1.65~~( 0.64) &  4~~ && Trumpler~16   & CPD -59$^\circ$2627 (O9 III,s), CPD -59$^\circ$2634 (O 9.5V,t)\\
13 & 10:45:14.1 & -59:57:53 && 2m &     81 &     75 & ~~1.64~~( 0.37) &  4~~ && {\em Spitzer} F  $\star$ & HD 305533 (B0.5 Vnn (shell):,u) \\
14 & 10:45:34.0 & -59:48:33 && 2    &     41 &     41 & ~~2.14~~( 0.78) & 10~~~&& {\em Spitzer} G        & FO 15 (O5.5Vz + O9.5 V,v) \\
15 & 10:45:36.6 & -60:12:50 && 1    &     52 &     32 & ~~1.82~~( 0.43) &  6~~ && {\em Spitzer} H  $\star$ & \nodata \\
16 & 10:45:47.1 & -60:05:07 && 2    &     31 &     31 & ~~1.62~~( 0.96) &  4~~ && {\em Spitzer} K  $\star$ & CPD -59$\circ$2660 (B0.5 V,w) \\
17 & 10:45:53.2 & -59:56:53 && 2    &     96 &     71 & ~~1.84~~( 0.63) &  6~~ && Treas Chest   & CPD -59$^\circ$2661 (O9.5 V,x)\\
     &                   &                     &&       &           &          &                              &     && {\em Spitzer} L $\star$ & \\
18 & 10:45:54.4 & -59:59:59 && 2    &     40 &     25 & ~~1.40~~( 0.20) &  2~~ && {\em Spitzer} J $\star$ & \nodata \\
19 & 10:45:59.5 & -60:08:32 && 1    &     25 &     23 & ~~2.05~~( 0.61) &  9~~ && {\em Spitzer} K-SE $\star$ & \nodata \\
20 & 10:47:14.7 & -60:05:48 && 1m &   137 &     92 & ~~1.72~~( 0.75) &  5~~ && Bochum~11 $\star$  & HD 93632 (O5 I-IIIf,y) \\
A  & \nodata       & \nodata    && 1    & 2867 & 2343 & ~~1.57~~( 0.58) &  3~~ && \nodata          & \nodata \\
B  & \nodata       & \nodata    && 1    & 1183 & 1116 & ~~1.67~~( 0.60) &  4~~ && \nodata           & \nodata \\
C  & \nodata       & \nodata    && 1   & 1407 & 1069 & ~~1.65~~( 0.63) &  4~~ && \nodata          & \nodata \\
D $\dagger$ & \nodata & \nodata  && 1   & 5271 & 3899 & ~~1.73~~( 0.85) &  5~~ && \nodata          & \nodata 
\enddata

\vspace*{-0.2in}
\tablecomments{ ~\\
$\star$ Cluster within the Collinder~228 and South Pillars region \\
$\dagger$ Region D contains stars not assigned to clusters 1$-$20 or Regions A$-$C}

\end{deluxetable}

\begin{deluxetable}{crrrrcrr}
\tablecaption{Small Groups of X-ray Stars Outside of Major Clusters}
\tabletypesize{\footnotesize}
\centering
\tablewidth{0pt}
\tablecolumns{8}
\tablehead{
\multicolumn{4}{c}{X-ray selected cluster} & \multicolumn{3}{c}{Absorption} & \colhead{{\em Spitzer}}  \\ 
\colhead{CCCP-Gp} & \multicolumn{2}{c}{RA ~(J2000)~Dec} & \colhead{$N$}
& \colhead{$N_{ME}$} & \colhead{$\langle MedE\rangle$} & \colhead{$A_V$} & \colhead{cluster} \\
 & & & & & keV & mag & \\
\colhead{(1)} & \colhead{(2)} & \colhead{(3)} & \colhead{(4)} & \colhead{(5)} & \colhead{(6)} & \colhead{(7)} & \colhead{(8)} 
}

\startdata
~1 & 10:42:13.5 & -59:35:59 &    5 &    5 & 2.32~~(1.24) &  11 & \nodata\\
~2 & 10:42:25.8 & -59:46:21 &    6 &    1 & \nodata & \nodata  & \nodata\\
~3 & 10:42:46.7 & -59:46:56 &    8 &    0 & \nodata & \nodata  & \nodata\\
~4 & 10:42:47.3 & -59:25:35 &   11 &    8 & 2.30~~(0.69) &  11 & \nodata\\
~5 & 10:42:49.5 & -59:09:05 &    6 &    2 & \nodata & \nodata  & \nodata\\
~6 & 10:42:52.8 & -60:13:20 &    7 &    3 & \nodata & \nodata  & \nodata\\
~7 & 10:42:53.4 & -59:26:15 &    7 &    5 & 1.89~~(0.74) &   7 & \nodata\\
~8 & 10:43:34.6 & -60:00:18 &    9 &    5 & 1.26~~(0.10) &   1 & A $\star$ \\
~9 & 10:43:41.0 & -60:01:16 &    8 &    3 & \nodata & \nodata  & A $\star$ \\
10 & 10:43:46.6 & -60:02:24 &    7 &    0 & \nodata & \nodata  & A $\star$ \\
11 & 10:43:59.9 & -60:01:59 &    6 &    2 & \nodata & \nodata  & A $\star$ \\
12 & 10:44:15.0 & -60:00:05 &    5 &    1 & \nodata & \nodata  & B $\star$ \\
13 & 10:44:20.8 & -59:59:03 &    9 &    9 & 1.39~~(0.18) &   2 & B $\star$ \\
14 & 10:44:22.5 & -59:59:37 &  10 &  10 & 1.16~~(0.16) &   1 & B $\star$ \\
15 & 10:44:22.6 & -59:25:14 &  12 &  11 & 1.52~~(0.22) &   3  & \nodata\\
16 & 10:44:26.3 & -59:59:55 &    7 &    7 & 1.34~~(0.19) &   2 & B $\star$ \\
17 & 10:44:40.2 & -59:46:52 &    9 &    9 & 1.55~~(0.53) &   3 & $\star$ \\
18 & 10:44:51.9 & -60:25:09 &    5 &    5 & 4.79~~(1.60) & 125 & \nodata\\
19 & 10:44:56.7 & -59:24:51 &  10 &    7 & 1.26~~(0.18) &   1 & \nodata\\
20 & 10:44:58.5 & -59:47:11 &    8 &    7 & 1.43~~(0.25) &   2 & $\star$ \\
21 & 10:45:30.0 & -59:57:41 &  12 &    7 & 1.27~~(0.16) &   1 & $\star$ \\
22 & 10:45:36.8 & -59:57:58 &  11 &    4 & 1.50~~(0.12) &   2 & $\star$ \\
23 & 10:45:38.5 & -60:00:44 &    7 &    7 & 1.47~~(0.38) &   2 & J $\star$ \\
24 & 10:45:40.4 & -60:01:07 &    8 &    8 & 1.78~~(1.19) &   6 & J $\star$ \\
25 & 10:45:44.9 & -59:55:27 &    7 &    4 & 1.57~~(0.10) &   3 & L $\star$ \\
26 & 10:46:05.8 & -59:56:55 &  11 &    7 & 1.93~~(0.90) &   7 & L $\star$ \\
27 & 10:46:06.0 & -59:58:59 &  12 &    4 & 1.70~~(0.68) &   4 & M $\star$ \\
28 & 10:46:50.7 & -60:04:23 &    9 &    6 & 1.44~~(0.23) &   2 & N $\star$ \\
29 & 10:46:52.0 & -60:06:03 &    8 &    5 & 1.61~~(0.35) &   4 & N $\star$ \\
30 & 10:46:54.4 & -60:04:39 &  11 &    7 & 1.29~~(0.25) &   1 & N $\star$ \\
31 & 10:47:17.4 & -60:07:51 &  10 &    4 & 1.43~~(0.24) &   2 & N $\star$ 
\enddata

\tablecomments{ ~\\
$\star$ Group within the Collinder~228 and South Pillars region }

\end{deluxetable}

\begin{deluxetable}{cccccrcccccc}
\tablecaption{Carina Cluster Membership} 
\rotate \hspace*{-0.5in}
\tabletypesize{\scriptsize}
\centering
\tablewidth{0pt}
\tablecolumns{12}
\tablehead{ \multicolumn{8}{c}{Source} && \multicolumn{3}{c}{Membership}\\
\cline{1-8} \cline{10-12}
\colhead{Seq} & \colhead{Label\tablenotemark{a}} & \colhead{CXOGNC} & \multicolumn{2}{c}{RA ~(J2000)~~Dec} & \colhead{NetCounts\_t}
& \colhead{$\log F_{t,photon}$} & \colhead{$MedE$\tablenotemark{b}} & & \colhead{Class} & \colhead{Cl/Gp} & \colhead{Reg} \\
\colhead{} & \colhead{} & \colhead{} & \multicolumn{2}{c}{deg} & \colhead{cnts}
& \colhead{photon~s$^{-1}$~cm$^{-2}$} & \colhead{keV} & & \colhead{} & \colhead{} & \colhead{}\\
\colhead{(1)} & \colhead{(2)} & \colhead{(3)} & \colhead{(4)} & \colhead{(5)} & \colhead{(6)}
& \colhead{(7)} & \colhead{(8)} & & \colhead{(9)} & \colhead{(10)} & \colhead{(11)}
}
\startdata
 8044 & CTr16\_1375 & 104454.19-594536.9 & 161.225804 & -59.760260 &    48.0 & -5.65 &   1.70 &  & H2 & C9 & B\\
 8045 &    C4\_1687 & 104454.20-592221.0 & 161.225861 & -59.372510 &     4.7 & -6.46 &   1.31 &  & H2 & C8 & A\\
 8046 &    SP1\_251 & 104454.21-601118.2 & 161.225878 & -60.188400 &     4.5 & -6.50 &   1.45 &  & H2 & \nodata & D\\
 8047 & CTr16\_1373 & 104454.21-594703.6 & 161.225883 & -59.784353 &     4.6 & -6.64 &   1.30 &  & H2 & \nodata & B\\
 8048 & CTr14\_3633 & 104454.22-593118.9 & 161.225922 & -59.521931 &    17.6 & -6.09 &   1.29 &  & H2 & \nodata & A\\
 8049 & CTr16\_1376 & 104454.23-594556.6 & 161.225985 & -59.765746 &     4.0 & -6.71 &   0.82 &  & H1 & C9 & B\\
 8050 &    C4\_1688 & 104454.28-591855.0 & 161.226194 & -59.315283 &     9.4 & -6.16 &   1.48 &  & H2 & \nodata & A\\
 8051 & CTr16\_1377 & 104454.29-594533.5 & 161.226232 & -59.759316 &     8.1 & -6.42 &   1.04 &  & H2 & C9 & B\\
 8052 &    SP1\_253 & 104454.31-601131.3 & 161.226313 & -60.192030 &     3.6 & -6.60 &    \nodata &  & H2 & \nodata & D\\
 8053 &    C4\_1690 & 104454.32-592011.2 & 161.226341 & -59.336467 &    69.8 & -5.29 &   1.51 &  & H2 & C8 & A\\
 8054 &    SP5\_384 & 104454.33-602545.2 & 161.226388 & -60.429224 &     4.6 & -5.71 &   2.55 &  & H4 & \nodata & D\\
 8055 & CTr16\_1378 & 104454.36-594355.3 & 161.226525 & -59.732032 &    10.0 & -6.33 &   1.16 &  & H2 & \nodata & B\\
 8056 &    SP5\_385 & 104454.37-602545.7 & 161.226543 & -60.429364 &     4.5 & -5.73 &   2.90 &  & H4 & \nodata & D\\
 8057 & CTr16\_1380 & 104454.38-593925.2 & 161.226611 & -59.657025 &    57.6 & -5.55 &   1.74 &  & H2 & \nodata & B\\
 8058 & CTr16\_1379 & 104454.38-594423.8 & 161.226624 & -59.739970 &    67.9 & -5.50 &   1.44 &  & H2 & \nodata & B\\
 8059 & CTr16\_1381 & 104454.40-594550.4 & 161.226693 & -59.764002 &     3.0 & -6.86 &    \nodata &  & H2 & C9 & B\\
 8060 & CTr16\_1382 & 104454.42-594527.8 & 161.226752 & -59.757737 &     5.2 & -6.59 &   1.61 &  & H2 & C9 & B\\
 8061 &    C4\_1691 & 104454.43-591928.1 & 161.226805 & -59.324479 &    10.6 & -6.10 &   1.34 &  & H2 & C8 & A\\
\enddata
 
\tablenotetext{a}{Source labels identify a CCCP pointing; they do not convey membership in astrophysical clusters.}

\tablenotetext{b}{This quantity is named MedianEnergy\_t in \citet{Broos10}.}

\tablecomments{The full table with 14,368 sources appears in the electronic edition of the Journal}
 
 \end{deluxetable}

\end{document}